\documentclass[a4paper,12pt]{article}
\pdfoutput=1 % if your are submitting a pdflatex (i.e. if you have
\usepackage{jheppub} % for details on the use of the package, please
\usepackage[T1]{fontenc} % if needed
\usepackage{enumitem} % To use alphabet or roman number instead of number in enumerate
\usepackage{verbatim}
\def\be{\begin{eqnarray}}
\def\ee{\end{eqnarray}}
\def\nn{\nonumber}

\title{\boldmath Quark generalized TMDs at skewness and Wigner Distribution in boost invariant longitudinal space}

\author[a,1]{Vikash Kumar Ojha,\note{Corresponding author.}}
\author[a]{Sujit Jana,}
\author[b]{Tanmay Maji}
\affiliation[a]{Sardar Vallabhbhai National Institute of Technology Surat,\\Icchanath, Gujarat, India}
\affiliation[b]{Indian Institute of Technology Hyderabad,\\Kandi, Telangana, India}
\emailAdd{vko@svnit.ac.in}
\emailAdd{d21ph010@phy.svnit.ac.in}
\emailAdd{tanmayhep@gmail.com}
\abstract{ The boost-invariant longitudinal space, defined by the parameter $\sigma = \frac{1}{2}b^- P^+$, can be studied from the Fourier transformation of distributions over the conjugate variable skewness $\xi$.  We investigate quark Wigner distributions in the $\sigma$ space in dressed quark model and found diffraction patterns that are analogous to the single slit experiment of light in optics. The width of the central maxima varies with energy transfer to the system and essentially $\xi$ behaves like a slit-width. Qualitatively similar diffraction pattern is reported recently in other models.  In this model, we compute all the leading twist GTMDs with non-zero skewness for quarks which provides Wigner distributions under Fourier transformation.  
}

\begin{document} 
\maketitle
\flushbottom

\section{Introduction}
The structure of the nucleon, one of the most significant active areas of current research, requires both the perturbative and the non-perturbative methods in Quantum Chromodynamics (QCD). The non-perturbative method usually uses the distributions function to describe the various aspect of nucleon structure. The parton distribution functions (PDFs) \cite{Collins:1981uw,Martin:1998sq,Gluck:1994uf,Gluck:1998xa} were first defined that provide information about the number density of partons carrying a particular amount of the longitudinal momentum fraction of nucleon. Generalized parton distributions (GPDs) \cite{Ji:1996nm,Diehl:2003ny,Belitsky:2005qn,Goeke:2001tz}, and transverse momentum dependent PDFs (TMDs) \cite{Mulders:1995dh,Barone:2001sp,Bacchetta:2006tn,Brodsky:2002cx,Bacchetta:2017gcc} were defined later to give a three-dimensional picture of the nucleon. The extraction and analysis of GPDs, TMDs are among the core goal of upcoming colliders, including the EIC \cite{Accardi:2012qut,Anderle:2021wcy,Boer:2011fh}. Both GPDs and TMDs can be considered as a special case of more generalized distributions GTMDs.

Generalized TMDs (GTMDs) are defined through the most general, off-diagonal, quark-quark correlator \cite{Meissner:2008ay,Meissner:2009ww,Lorce:2013pza}. It gained significant attention in recent times as it has direct model-independent relation to the orbital angular momentum and spin-orbit correlation of partons \cite{Bhattacharya:2017bvs,Hatta:2016dxp,Ji:2016jgn,Hatta:2016aoc,Bhattacharya:2022vvo,Hagiwara:2017fye,Zhou:2016rnt,Lorce:2014mxa,Echevarria:2016mrc,Bhattacharya:2018lgm}. It may even be possible to access the GTMDs in future experiments. GTMDs have been studied previously in different models, but mainly for the zero skewness. The kinematical variable skewness ($\xi$) measures the fraction of longitudinal momentum transferred to the target state. The value of $\xi$ lies between -1 and 1, and for processes in which momentum transfer is purely in the transverse direction, it is 0.  Most of the previous GTMDs and Wigner distribution studies have been done with the assumption that momentum transferred to the target state during the interaction is purely in a transverse direction. This assumption is chosen often as it dramatically simplifies the calculation involved. But if we want to access the GTMDs in the experiments, we must choose the skewness as non-zero for our analysis because the skewness is rarely zero in experiments. This makes it an interesting case to study the GTMDs at non-zero skewness. In this article, We obtain the GTMDs of quarks in the light-front dressed quark model for non-zero skewness of the target.

Along with GTMDs, the Wigner distribution \cite{Wigner:1932eb,Ji:2003ak} is also at the top of the
distribution functions hierarchy related to partons. The Wigner distribution are the Fourier transform of the GTMDs and have been studied extensively in recent times \cite{Mukherjee:2014nya,Mukherjee:2015aja,Lorce:2011kd,Lorce:2011ni,Kaur:2018dns,Liu:2015eqa,Maji:2022tog,Chakrabarti:2016yuw,Chakrabarti:2017teq,More:2017zqp,More:2017zqq}. 
The 6-dimensional Wigner distribution has been proposed
recently as the relativistic version of the more popular 5-dimensional Wigner distribution \cite{Han:2022tlh}.

The boost-invariant longitudinal spatial coordinate ($\sigma$), also called impact parameter in longitudinal space, was introduced in \cite{Brodsky:2006in} as conjugate to the longitudinal momentum transfer to the target. The GPDs were explored in the $\sigma-$space as it
gives a complete 3-D picture of hadron when studied along with the distribution in the impact parameter space\cite{Brodsky:2006ku,Chakrabarti:2005zm,Chakrabarti:2015ama,Manohar:2010zm,Mukherjee:2011an}.
Recently Wigner distribution in $\sigma$-space was studied for quark in the light-front quark-diquark model \cite{Maji:2022tog}. 
This article analyzes the Wigner distribution in the $\sigma$ space for quark in the light-front dressed quark model.

The article is organized as follows. In Section 2, we fix the target state's notation, conventions, and kinematics. In section 3, we briefly summarize the concept and mathematical tools of the light-front dressed quark model required for deriving the analytical expression of GTMDs and Wigner distributions. Section 4.1 presents the analytical results of GTMDs obtained in the light-front dressed quark model. In section 4.2, we rewrite the GTMDs in a modified form, making it easier to compare with the GTMDs obtained earlier for zero skewness. The GTMDs in the
limit of zero skewness is presented in section 4.3. In section 4.4, we discuss the spin and OAM of quark and its relation to GTMDs. Section 5.1 defines the Wigner distribution for different polarization of quark and target state. Section 5.2 expresses the Wigner distribution as the Fourier transform of the GTMDs. Section
5.3 consists of the plots of Wigner distributions in the boost-invariant longitudinal
space ($\sigma$). Finally, we conclude in section 6.

%%%%%%%%%%%%%%%%%%%%%%%%%%% New Section %%%%%%%%%%%%%%%%%%%%%%%%%%%%%%

\section{Kinematics \label{sec:Conven}}
      We choose to work in light-front coordinate system $(x^+, x^-, x_\perp)$ where the light-front time and longitudinal space variable are defined as $x^{\pm} = x^0 \pm x^3 $ and all the other conventions are listed in \cite{Zhang:1994ti,Harindranath:1996hq}.
We consider a system of a quark dressed by gluons interacting with a virtual photon probing with energy transfer $t=\Delta^2$ into the system. The longitudinal momentum transfer is defined as $\xi = \Delta^+/2P^+$. The initial  dressed quark is labeled by momentum $p$ and the final momentum of dressed quark system is labeled by $p^\prime$  can be expressed in the  symmetric frame \cite{Brodsky:2000xy} for kinematics as
\be
    p &=&\Bigg((1+\xi)P^+,\Delta_\perp/2,\frac{m^2+\Delta_\perp^2/4}{(1+\xi)P^+}\Bigg) ;\\
    p'&=&\Bigg((1-\xi)P^+,-\Delta_\perp/2,\frac{m^2+\Delta_\perp^2/4}{(1-\xi)P^+}\Bigg),
\ee
where, the average momentum of the system $P=(p + p^\prime)/2$. 
The four-momentum transfer from the target state is
\be
    \Delta = p'-p = \Bigg(2\xi P^+, \Delta_\perp, \frac{t+\Delta_\perp^2}{2\xi P^+} \Bigg),
\ee
where,
\be
    t=-\frac{4\xi^2 m^2 + \Delta_\perp^2}{1-\xi^2}
\ee
can be easily derived from constraint $\Delta^-=p^{\prime-}-p^-$ provided by energy momentum conservation. The mass of the system is represented as $m$. The struck quark carries a momentum fraction $x=k^+/p^+$ of the system and the quark four-momentum can be written as 
\be 
k\equiv \Big(x P^+, k_\perp, k^- \Big)\label{mom_k}.
\ee

%
%

%%%%%%%%%%%%%%%%%%%%%%%%%%%% New Section %%%%%%%%%%%%%%%%%%%%%%%%%%%%%%%

\section{Light-front Dressed Quark Model}
Proton is a highly complex object at a partonic scale as it is a bound state of 3 valence
quarks and gluons. Naturally, it is challenging to analyze such a multi-particle bound
state. Some simplified model of the bound state of quarks has been used to study the bound state of partons, like the quark-diquark model \cite{Kaur:2018dns,Kumar:2015coa,Brodsky:2000xy,Bacchetta:2008af,Bacchetta:2015qka}, chiral quark soliton model \cite{Lorce:2011ni,Lorce:2011kd}, AdS/QCD quark–diquark model \cite{Chakrabarti:2017teq,Chakrabarti:2016yuw,Maji:2016yqo},
and dressed quark model \cite{Harindranath:1998pc,Harindranath:1998ve,Zhang:1993dd}. The dressed quark model has significance as it contains a gluonic degree of freedom which allows studying the behavior of gluons in a bound state. We use dressed quark model to analyze the GTMDs and Wigner
distribution in boost-invariant longitudinal space. This article considers only
the GTMDs and Wigner distributions of quarks. A dressed quark can be considered a bound state of a quark and a gluon. We represent a dressed quark state with momentum $p$ and helicity $\sigma$ as \cite{Zhang:1993dd}
\be \label{fockse}
  \Big{| }p^{+},p_{\perp},\sigma  \Big{\rangle} = \Phi^{\sigma}(p) b^{\dagger}_{\sigma}(p)
 | 0 \rangle +
 \sum_{\sigma_1 \sigma_2} \int [dp_1]
 \int [dp_2] \sqrt{16 \pi^3 p^{+}}
 \delta^3(p-p_1-p_2) \nn \\ \Phi^{\sigma}_{\sigma_1 \sigma_2}(p;p_1,p_2) 
b^{\dagger}_{\sigma_1}(p_1) 
 a^{\dagger}_{\sigma_2}(p_2)  | 0 \rangle;
\ee
where $[dp] =  \frac{dp^{+}d^{2}p_{\perp}}{ \sqrt{16 \pi^3 p^{+}}}$. $b^{\dagger}$ and $a^{\dagger}$ are creation operators 
for quark and gluon respectively. $ \Phi^{\sigma}(p)$ represents single particle wave function which contributes only when
$x=1$. $ \Phi^{\sigma}_{\sigma_1 \sigma_2}$ is the two-particle wave function and its amplitude gives the probability to find a 
bare quark with momentum $p_1$ and helicity $\sigma_1$ and a bare gluon with momentum $p_2$ and helicity $\sigma_2$ inside the
target state. Both $ \Phi^{\sigma}(p)$ and $ \Phi^{\sigma}_{\sigma_1 \sigma_2}$ 
can be obtained using light-front eigenvalue equation in the Hamiltonian approach. The boost invariant LFWF  
and two-particle LFWF are related by the relation $\Psi^{\sigma}_{\sigma_1
\sigma_2}(x, q_\perp) =   
\Phi^{\sigma}_{\sigma_1 \sigma_2}
\sqrt{P^+}$.  Here we have used the Jacobi momenta $(x_i, q_{i \perp})$ : 
\be \label{jacmom}
p_i^+= x_i p^+, ~~~~~~~~~~q_{i \perp}= k_{i \perp}+x_i p_\perp
\ee
so that $\sum_i x_i=1, \sum_i q_{i\perp}=0$. The expression for two-particle LFWFs can be calculated 
perturbatively as \cite{PhysRevD.59.116013}:

\be \label{tpaq}
\Psi^{\sigma a}_{\sigma_1 \sigma_2}(x,q_{\perp}) = 
\frac{1}{\Big[    m^2 - \frac{m^2 + (q_{\perp})^2 }{x} - \frac{(q_{\perp})^2}{1-x} \Big]}
\frac{g}{\sqrt{2(2\pi)^3}} T^a \chi^{\dagger}_{\sigma_1} \frac{1}{\sqrt{1-x}}
\nn \\ \Big[ 
-2\frac{q_{\perp}}{1-x}   -  \frac{(\sigma_{\perp}.q_{\perp})\sigma_{\perp}}{x}
+\frac{im\sigma_{\perp}(1-x)}{x}\Big]
\chi_\sigma (\epsilon_{\perp \sigma_2})^{*}.
\ee
We are using the two component formalism \cite{Zhang:1993dd}, where $\chi$ is the two
component spinor, $T^a$ are the color $SU(3)$ matrices, $m$ is 
the mass of the quark, $\sigma_{\perp}$ $(\perp = 1,2)$ is Pauli matrices and $\epsilon_{\perp \sigma_2}$ is the polarization
vector of the gluon. 
The light-front wavefunctions for quark are expressed as the function Jacobi momenta $(x',q'_\perp)$. We choose $(x',q'_\perp)$ and $(y,q_\perp)$ as the initial and final Jocbi momenta respectively for dressed quarks. 

%%%%%%%%%%%%%%%%%%%%%%%%% New Section %%%%%%%%%%%%%%%%%%%%%%%%%%%%%%%%%%%
\section{Generalized TMDs of quark}
%

% \section{Kinematics}
In the light-front gauge $z^+=0$, the quark-quark correlator for GTMDs
$W^{[\Gamma]}_{\lambda,\lambda^\prime} (x,\xi,{\bf k_{\perp}},{\bf\Delta_{\perp}}) $ is defined through the non-diagonal matrix element of the bi-local quark field \cite{Meissner:2009ww} as
\be \label{qqc}
W^{[\Gamma]}_{\lambda,\lambda^\prime} (x,\xi,{\bf k_{\perp}},{\bf\Delta_{\perp}}) %\Big{\langle }p^{+},\frac{\Delta_{\perp}}{2},\sigma  \Big{|}
%W^{[\Gamma]} (0_{\perp},k_{\perp},x)  \Big{|}  p^{+},-\frac{\Delta_{\perp}}{2},\sigma \Big{\rangle } 
% \nn \\ \nn \\
=\frac{1}{2}\int\frac{dz^{-}}{2\pi}\frac{d^{2} z_{\perp}}{(2\pi)^2}e^{ip\cdot z}
 \Big{\langle } p^{\prime},\lambda^{\prime} \Big{|}
\overline{\psi}(-z/2) \mathcal{W}_{[-z/2,z/2]}\Gamma \psi(z/2) \Big{|}
p,\lambda \Big{\rangle }
\Big{|}_{z^{+}=0},\nn
\\
\ee 
where, the ${|}p,\lambda {\rangle } $ and ${|} p^{\prime},\lambda^{\prime}\rangle $ represents the initial and final state of the dressed quark system. The Wilson line  $\mathcal{W}_{[-z/2,z/2]}$ is defined as a gauge link between the two quark fields $\psi(z/2)$ and $\bar{\psi}(-z/2)$ situated at two different points. Here, we the Wilson is reduced to unity in the light-front gauge. In the both side of the correlator, $\Gamma$ stands for the polarization structure and at leading twist it is an element of the set $\{\gamma^+,\gamma^+\gamma^5,i\sigma^{+j}\gamma^5\}$ corresponding to the unpolarized, longitudinally polarized and transversely polarized cases respectively.
All GTMDs are the function of $(x,\xi,{\bf k^2_\perp},{\bf\Delta^2_\perp},{\bf k_\perp\cdot\Delta_\perp})$. However, we suppressed these arguments while writing the final expression of GTMDs. So, when we are writing $F_{1,1}$ in the following sections, it must be understood as $F_{1,1}(x,\xi,{\bf k^2_\perp},{\bf\Delta^2_\perp},{\bf k_\perp\cdot\Delta_\perp})$.
Using the dressed quark state from Eq.(\ref{fockse}) and the particle sector of quark fields $\psi(\pm z/2)$, one can express the correlator in terms of the overlap representation of light-front wave functions defined in Eq.(\ref{tpaq}).  
The quark-quark correlator for unpolarized, longitudinally polarized and transversely polarized quarks read as
\be
    W^{[\gamma^+]}_{\lambda,\lambda^\prime} (x,\xi,{\bf k_{\perp}},{\bf\Delta_{\perp}}) = 
    \sum_{\sigma_1, \sigma_{2},\lambda_{1}} 
    \Psi^{*\lambda'}_{\lambda_{1} \sigma_2}(x',q'^{\perp}) \chi_{\lambda_1}^{\dagger} \chi_{\sigma_1}
    \Psi^{\lambda}_{\sigma_1 \sigma_2}(y,q^{\perp}) \label{qqc-unpolarized}\\
    W^{[\gamma^+\gamma^5]}_{\lambda,\lambda^\prime} (x,\xi,{\bf k_{\perp}},{\bf\Delta_{\perp}}) = 
    \sum_{\sigma_1, \sigma_{2},\lambda_{1}} 
    \Psi^{*\lambda'}_{\lambda_{1} \sigma_2}(x',q'^{\perp}) \chi_{\lambda_1}^{\dagger}\sigma_3 \chi_{\sigma_1}
    \Psi^{\lambda}_{\sigma_1 \sigma_2}(y,q^{\perp}) \label{qqc-long-polarized}\\
    W^{[i\sigma^{j+}\gamma^5]}_{\lambda,\lambda^\prime} (x,\xi,{\bf k_{\perp}},{\bf\Delta_{\perp}}) = 
    \sum_{\sigma_1, \sigma_{2},\lambda_{1}} 
    \Psi^{*\lambda'}_{\lambda_{1} \sigma_2}(x',q'^{\perp}) \chi_{\lambda_1}^{\dagger} \sigma_j\chi_{\sigma_1}
    \Psi^{\lambda}_{\sigma_1 \sigma_2}(y,q^{\perp})\label{qqc-trans-polarized}\ee
where, initial (final) struck quark carries longitudinal momentum fraction $y (x^\prime)$ and transverse momentum $q_\perp (q_\perp^\prime)$ and the chosen kinematics are parameterized as 
\be
    x'&=\frac{x-\xi}{1-\xi} ~~,~~~
    q'_\perp=k_\perp-\frac{(1-x)}{(1-\xi)}\frac{\Delta_\perp}{2}\nn\\
    y&=\frac{x+\xi}{1+\xi} ~~,~~~
    q_\perp=k_\perp+\frac{(1-x)}{(1+\xi)}\frac{\Delta_\perp}{2}\nn
\ee
where $x$ is the momentum fraction corresponding to the quark average momentum $k$ as defined in Eq.(\ref{mom_k}). 
The billinear decomposition of the quark-quark correlator of Eq.(\ref{qqc}), at the leading twist, leads to the sixteen GTMDs \cite{Meissner:2009ww} and the explicit expansion is listed in Appendix-A for completeness. 
We derived the analytical expression for all the 16 GTMDs using Eq.(\ref{qqc-unpolarized}-\ref{qqc-trans-polarized}), and Eq.(\ref{Bilinear-Decomposition-F}-\ref{Bilinear-Decomposition-H}) and 
 the analytical results in this dressed quark model is as follows:\\
% \begin{enumerate}[label=(\alph*)]
   (a) For unpolarized quark, 
    \begin{align}
  F_{1,1}=&\frac{\alpha(1-\xi^2)}{2(1-x)^3}\Big[4(1+x^2-(3+x^2)\xi^2+2\xi^4)k^2_\perp+4m^2(1-x)^4\nn\label{f11}\\
  &+ 4(1-x)\xi(1+x^2-2\xi^2)k_\perp\cdot\Delta_\perp-(1-x)^2(1+x^2-2\xi^2)\Delta^2_\perp\Big]\\ 
  F_{1,2}=&-\beta\Big[((1+x)\xi(k_2\Delta_1-k_1\Delta_2)-2m^2x(1-x))\Delta^2_\perp \nn\\
  &+4m^2\xi(1+x)k_\perp\cdot\Delta_\perp\Big]\label{f12}\\
  F_{1,3}=&\frac{\beta}{4(1-x)^2}\Big[(k_1\Delta_2-k_2\Delta_1)((1+x^2-2\xi^2)(4(-1+\xi^2)k^2_\perp \nn\\
  &+(-1+x)^2\Delta^2_\perp)-8(-1+\xi^2)(-1+x)\xi k_\perp\cdot\Delta_\perp -4m^2(-1+x)^4)\nn\\
  &+8m^2(-1+x)^2(2(1+x)\xi k_\perp+(-1+x)x\Delta_\perp)\cdot k_\perp\Big]\\
  F_{1,4}=&\frac{\alpha}{(1-x)}\Big[2m^2(1+x)(1-\xi^2)\Big]
  \label{f14}
    \end{align}
 (b) For Longitudinally Polarized quark,
  \begin{align}
        G_{1,1}=&\frac{\alpha}{(1-x)}\Big[-2m^2(1+x)(1-\xi^2)\Big]\label{g11}\\
        G_{1,2}=&\beta\Big[2m^2\big((1-x)\xi\Delta^2_\perp-2(x+\xi^2)k_\perp\cdot\Delta_\perp\big)-(1+x)(k_2\Delta_1-k_1\Delta_2)\Delta^2_\perp\Big]\\
        G_{1,3}=&\frac{\beta}{4(1-x)^2}\Big[(k_2\Delta_1-k_1\Delta_2)(\xi(1+x^2-2\xi^2)(4(1-\xi^2)k^2_\perp-(1-x)^2\Delta^2_\perp)\nn\\
         &+ 4(1-\xi^2)(1-x)(1-x^2+2\xi^2)k_\perp\cdot\Delta_\perp)+4m^2(1-x)^2(4(x+\xi^2)k^2_\perp\nn\\
        &- \xi(1-x)(2k_\perp\cdot\Delta_\perp+(1-x)(k_2\Delta_1-k_1\Delta_2)))\Big]\\
        G_{1,4}=&\frac{\alpha(1-\xi^2)}{4(1-x)^3}\Big[(1+x^2-2\xi^2)(4(1-\xi^2)k^2_\perp+(1-x)(4\xi
        k_\perp-(1-x)\Delta_\perp)\cdot\Delta_\perp)\nn\\
        &-4m^2(1-x)^4\Big]
  \end{align}
(c) For transversely Polarized quark,
  \begin{align}
        H_{1,1}= &\beta[2m^2(1-\xi^2)(4\xi k_\perp\cdot\Delta_\perp-(1-x)\Delta^2_\perp)] \label{h11}\\
        H_{1,2}= &-\beta[2m^2(1-\xi^2)(4\xi k^2_\perp-(1-x)k_\perp\cdot\Delta_\perp)] \\
        H_{1,3}= &\frac{\alpha}{(1-x)^3k_\perp\cdot\Delta_\perp}\Big[4(x-\xi^2)(1-\xi^2)k^4_\perp+ 2(1-x)^2\xi(3+\xi^2)(k_1\Delta_2-k_2\Delta_1)k_\perp\cdot\Delta_\perp\nn\\ &
        +(1-x)(x-\xi^2)(4\xi(k^2_1\Delta^2_1+k^2_2\Delta^2_2)-(1-x)\Delta^2_\perp(k_\perp\cdot\Delta_\perp))+\xi(1-x)(4(1-x)\nn\\
        &(k^2_1\Delta^2_2+k^2_2\Delta^2_1)+\xi(k_1\Delta_2-k_2\Delta_1))-8(1-x)(1-2x+\xi^2)\xi k_1k_2\Delta_1\Delta_2\Big] 
         \end{align}
  \begin{align}
        H_{1,4}= &\frac{\beta}{k_\perp\cdot\Delta_\perp}\Big[m^2\xi\Delta^2_\perp(2(1+\xi^2)k_\perp\cdot\Delta_\perp)+4(k_1\Delta_2-k_2\Delta_1)-(1-x)\xi\Delta^2_\perp\Big]\\
        H_{1,5}= &-\beta\Big[m^2(2\xi(3+\xi^2)k_\perp\cdot\Delta_\perp +8\xi(k_1\Delta_2-k_2\Delta_1)-(1-x)(1+\xi^2)\Delta^2_\perp\Big] \\
        H_{1,6}= & \frac{\beta}{k_\perp\cdot\Delta_\perp}\Big[m^2(4\xi k^2_\perp(k_\perp\cdot\Delta_\perp-(k_2\Delta_1-k_1\Delta_2))-(1-x) (k^2_1(\Delta^2_1+\xi^2\Delta^2_2)\nn \\
        &+k^2_2(\xi^2\Delta^2_1+\Delta^2_2))-(1-x)(1-\xi^2)k_1k_2\Delta_1\Delta_2)\Big] \\
        H_{1,7}= &-\beta\Big[m^2(1-\xi^2)\Big(2(1+\xi^2)k_\perp\cdot\Delta_\perp-(1-x)\xi\Delta^2_\perp\Big)\Big]\\
        H_{1,8}= &\beta\Big[m^2(1-\xi^2)\Big(2(1+\xi^2)k^2_\perp-(1-x)\xi\Delta^2_\perp\Big)\Big] \label{h18}
    \end{align}
% \end{enumerate}    
%\subsection{Analytical Results for GTMDs}
 We define the function $D(k_\perp,x)$, and $\alpha(x,\xi, k^2_\perp,\Delta^2_\perp, k_\perp\cdot\Delta_\perp)$ as 
\begin{align}
    D(k_\perp,x)=&\Bigg(m^2-\frac{m^2+k_\perp^2}{x}-\frac{k^2_\perp}{1-x}\Bigg) \nn\\
   \text{and,~~} \alpha(x,\xi, k^2_\perp,\Delta^2_\perp,& k_\perp\cdot\Delta_\perp)=\frac{N}{D(q_\perp,y)D^*(q'_\perp,x')(x^2-\xi^2)}
    \end{align}
    %
  %  \begin{align}
   % \text{where,~~}
 %   N=\frac{g^2 C_f}{2(2\pi)^3},~~~ %D(q_\perp,y)=&\Bigg(m^2-\frac{m^2+q_\perp^2}{y}-\frac{(q_\perp)^2}{1-y}\Bigg) \nn\\
   % =&\frac{m^2 (x-1)^2+(\xi +1)^2 (q_1^2+q_2^2)}{(x-1) (x+\xi )}\\
%
%\text{and}~~~
%    D(q_\perp^\prime,x^\prime)=&\Bigg(m^2-\frac{m^2+(q_\perp^\prime)^2}{x^\prime}-\frac{(q_\perp^{\prime})^2}{1-x^\prime}\Bigg) \nn\\
%    =&\frac{m^2 (x-1)^2+(\xi -1)^2 ({q_1^\prime}^2+{q_2^\prime}^2)}{(x-1) (x-\xi )}.
%\end{align}
where $N=\frac{g^2 C_f}{2(2\pi)^3}$ with the strong coupling constant  $g$  and color $C_f$.
We again suppress the arguments of $\alpha(x,\xi, k^2_\perp,\Delta^2_\perp, k_\perp\cdot\Delta_\perp)$ and write it simply as $\alpha$. Introducing another function $\beta$ as
\be
\beta \equiv \frac{\alpha}{(1-x)(k_2\Delta_1-k_1\Delta_2)},
\ee
%  \subsection{Rewriting the Analytical Results for GTMDs}
  We express the form of GTMDs in a more convenient form, in terms of $\beta$,  that will be helpful to compare with the GTMDs results of the limit $\xi\rightarrow 0$. One can also rewrite the GTMDs results by separating the vanishable term at the zero skewness limit as
  \begin{align}
       \chi_{1,j}(x,\xi,{\bf k^2_\perp},{\bf\Delta^2_\perp},{\bf k_\perp\cdot\Delta_\perp})&=\hat{\chi}_{1,j}(x,\xi,{\bf k^2_\perp},{\bf\Delta^2_\perp},{\bf k_\perp\cdot\Delta_\perp})+\xi f_{1,j}(x,\xi,{\bf k^2_\perp},{\bf\Delta^2_\perp},{\bf k_\perp\cdot\Delta_\perp}) \nn\\
  \end{align}
  where, $\chi_{1,j} \equiv \{F_{1,j},~G_{1,j}, H_{1,j}\} $ with $j=1,\cdots,4$ for $F_{1,j},~G_{1,j}$ and $j=1,\cdots, 8$ for $H_{1,j}$. For example, $F_{1,4}$ can be written as
  \begin{align}
    F_{1,4}(x,\xi,{\bf k^2_\perp},{\bf\Delta^2_\perp},{\bf k_\perp\cdot\Delta_\perp})&=\mathcal{F}_{1,4}(x,\xi,{\bf k^2_\perp},{\bf\Delta^2_\perp},{\bf k_\perp\cdot\Delta_\perp})+\xi f_{1,4}(x,\xi,{\bf k^2_\perp},{\bf\Delta^2_\perp},{\bf k_\perp\cdot\Delta_\perp})
  \end{align}
  \begin{align}
    {\rm where},~~    \mathcal{F}_{1,4}&=\frac{\alpha}{(1-x)}[2m^2(1+x)], \nn\\
      \textrm{and~~} f_{1,4}=&\frac{\alpha}{(1-x)}\Big[-2m^2(1+x)\xi\Big].
  \end{align}

  In the limit $\xi\rightarrow 0$, the second term ($f_{1,4}$) vanishes, and in the first term, $\xi$ dependence comes only from the factor $D(q_\perp,y)D^*(q'_\perp,x')(x^2-\xi^2)$ in the denominator (absorbed in $\alpha$), which reduces to $D(q_\perp)D^*(q'_\perp)x^2$. 

  \begin{figure}[!htp]
\begin{minipage}[c]{1\textwidth}
\small{(a)}\includegraphics[width=7.8cm,height=6cm,clip]{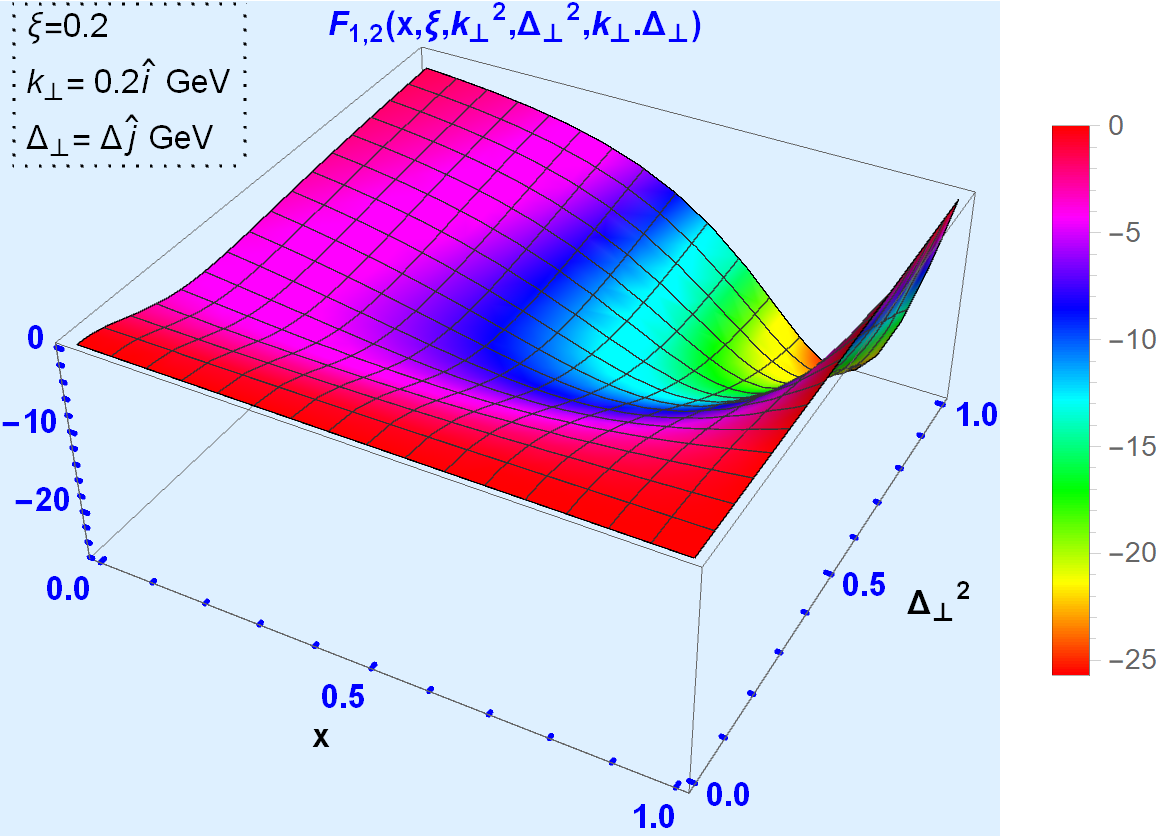}
\hspace{0.1cm}
\small{(b)}\includegraphics[width=7.8cm,height=6cm,clip]{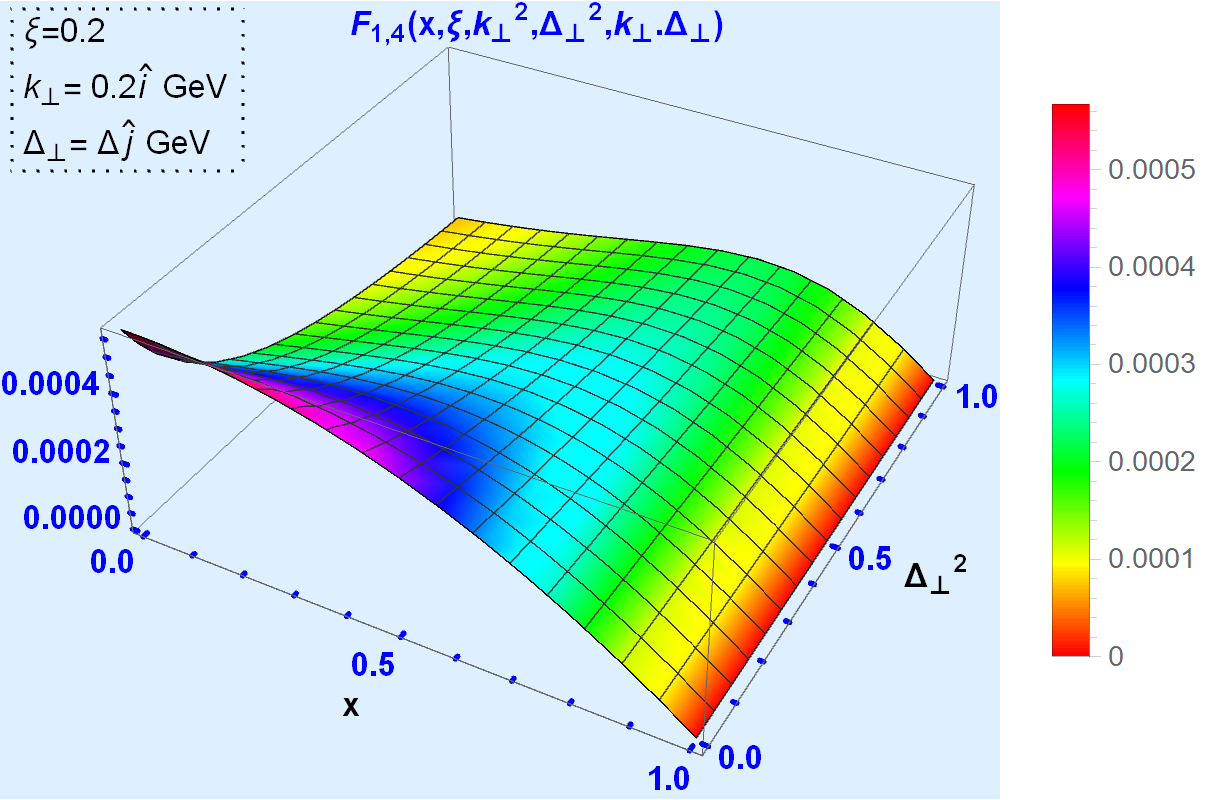} \\
\end{minipage}
\caption{\label{fig_x_Dp}
GTMDs (a) $F_{1,2}$, and (b) $F_{1,4}$ as function of $x$ and $\Delta^2_\perp$ for fixed values of $\xi=0.2$ and $k_\perp = 0.2 \hat{i}$ GeV. }
\end{figure}

To illustrate the numerical results for non-zero skewness, we concentrate on those GTMDs which have some physical implications at certain limit e.g., $F_{1,2}, F_{1,4}, G_{1,1}$ and $H_{1,1}$. The three-dimensional variation in the $x, \Delta^2_\perp$ plain of $F_{1,2}, F_{1,4}$ GTMDs for unpolarized dressed quark system are shown in the Fig.\ref{fig_x_Dp}(a) and (b) respectively. The transverse momentum of the struck quark is fixed along $x$-axis ($k_\perp=0.2 \hat{i}$ GeV) and $\Delta_\perp$ is chosen along the $y$-axis. This particular choices of transverse directions are made aiming to avoid contribution from the $k_\perp.\Delta_\perp$ terms in Eq.(\ref{f11}). We also fixed the longitudinal momentum transfer at $\xi=0.2$, and choose the mass of bare quark to be $3.3$ MeV. We noticed that, the distribution is negative and the peak shifts towards higher value of $x$ with the increasing $\Delta^2_\perp$.   At the TMDs limit, $\Delta_\perp =0, \xi=0$,  the GTMDs $F_{1,2}$ reduces to the Sivers TMDs that plays crustal role in the spin-transverse momentum correlation which leads to the breaking of axial symmetry and provide left-right axial shifting in the transverse momentum plain.  
The Fig.\ref{fig_x_Dp}(b) shows the three-dimensional variation for $ F_{1,4}$ GTMD. The distribution is positive and becomes flatten at higher values of $\Delta^2_\perp $. At the limit $\xi=0$, $ F_{1,4}$ contributes to the correlation between quark orbital angular momentum(OAM) and spin of the system given by 
\be
l^{q}_{z} = -\int dx d^{2}k_{\perp} \frac{k_{\perp}^2}{m^2} F_{14}.
\ee
If $l^{q}_{z} > 0 $, quark OAM tends to align along the spin of the system and for  $l^{q}_{z} < 0 $ they are tending to anti-align. In our model the  $l^{q}_{z} = -0.125074$ for $Q=5$ GeV, $Q$ is large scale involved in the process, which indicates that the quark OAM is anti-align to the spin of the dressed quark system. 

  \begin{figure}[!htp]
\begin{minipage}[c]{1\textwidth}
\small{(a)}\includegraphics[width=7.8cm,height=6cm,clip]{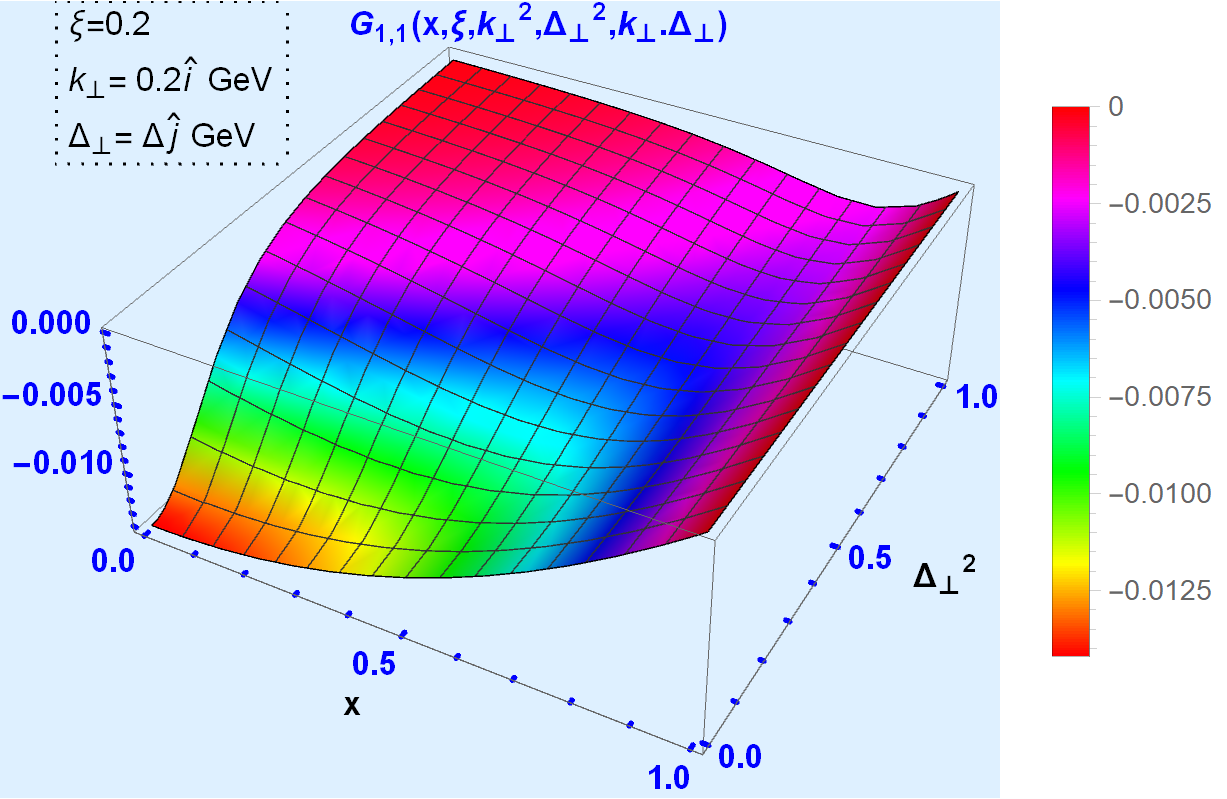}
\hspace{0.1cm}
\small{(b)}\includegraphics[width=7.8cm,height=6cm,clip]{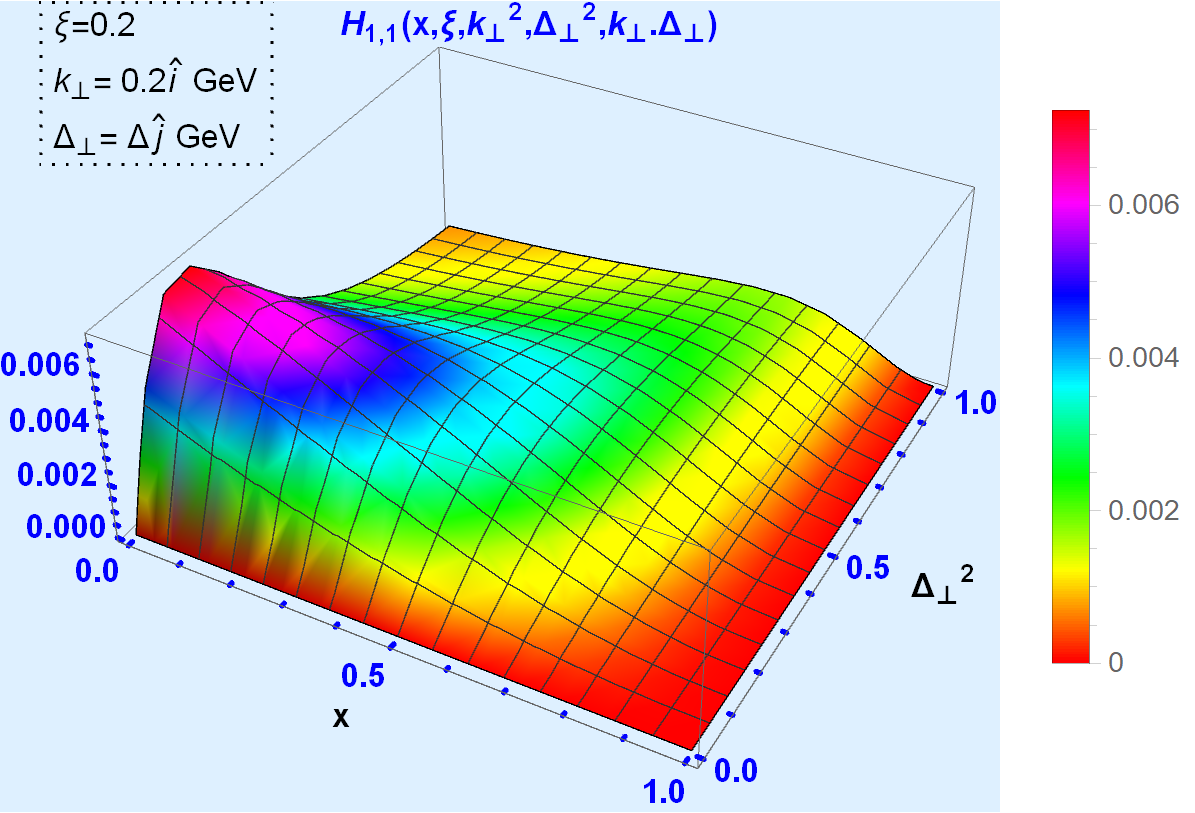} \\
\end{minipage}
\caption{\label{fig_x_Dp2}
GTMDs (a) $G_{1,1}$, and (b) $H_{1,1}$ as function of $x$ and $\Delta^2_\perp$ for fixed values of $\xi=0.2$ and $k_\perp = 0.2 \hat{i}$ GeV.}
\end{figure}
Fig.\ref{fig_x_Dp2}(a) shows the $x, \Delta^2_\perp$ variation of $G_{1,1}$ at $\xi=0.2$, $k_\perp = 0.2 \hat{j}$ GeV. This is one of the GTMDs at leading twist corresponding to the longitudinally polarized dressed quark system. Again the peak of the distribution goes to the large $x$ region for higher value of $ \Delta^2_\perp$.  Comparing the analytical results of Eqs.(\ref{f14}) and Eqs.(\ref{g11}), we find that the relation $F_{1,4}= -G_{1,1}$ is still valid  for non-vanishing skewness in other words the model dependent result $F_{1,4}= -G_{1,1}$ is also satisfied even for the finite longitudinal momentum transfer to the process  \cite{Mukherjee:2014nya}. The GTMD $G_{1,1}$ also provides the correlation between quark spin and OAM by 
\be
C^{q}_{z} = \int dx d^{2}k_{\perp} \frac{k_{\perp}^2}{m^2} G_{11},
\ee
at zero skewness limit. For $C^{q}_{z}>0$ quark spin and OAM lends to align and they are anti-align for $C^{q}_{z}<0$. In this model we have found a anti-parallel correlation between quark spin and OAM with strength $C^{q}_{z}= -0.125074$ for $Q=5$ GeV. 
$H_{1,1}$ is one of the GTMDs found for transversely polarized dressed quark system and is shown in Fig.\ref{fig_x_Dp2}(b) for $\xi=0.2$, $k_\perp = 0.2 \hat{j}$ GeV and $\Delta_\perp$ is along $\hat{y}$. For this transverse axise choices the term containing the dot product $k_\perp.\Delta_\perp$ will vanish in Eq.(\ref{h11}). This kind of terms leads to a dipoler distribution in the transverse momentum plain as shown in the \cite{More:2017zqq} for the $\xi=0$ limit. The TMD limit of this distribution provides the Boer-Mulders TMDs.

  %%%%%%%%%%%%% New SubSection %%%%%%%%%%%%%%%%%
  
  \subsection{GTMDs in limit $\xi\rightarrow 0$}
  GTMDs of unpolarized and longitudinally polarized quark in the dressed quark model has already been calculated for $\xi= 0$ limit \cite{Mukherjee:2014nya} and it is important to obtain the GTMDs in
the limit $\xi\rightarrow 0$ to ensure or check the results' correctness. Our results agree with the results of \cite{Mukherjee:2014nya} for unpolarized and longitudinally polarized case ($F_{1,i},G_{1,i}$). There are no previous results for transversely polarized quarks ($H_{1,i}$) for $\xi\rightarrow 0$ limit in this model. Thus for completeness, we list up the results of the GTMDs $H_{1,i}$ corresponding to the transversely polarized dressed quark system at $\xi\rightarrow 0$ limit as   
   \begin{align}
    H_{1,1}&=\frac{2m^2N\Delta^2_\perp}{D^*(q'_\perp)D(q_\perp)x^2(k_2\Delta_1-k_1\Delta_2)},\\
    H_{1,2}&=\frac{2m^2Nk_\perp\cdot\Delta_\perp}{D^*(q'_\perp)D(q_\perp)x^2(k_2\Delta_1-k_1\Delta_2)},\\
    H_{1,3}&=\frac{N(4k^2_\perp-(1-x)^2\Delta^2_\perp)}{D^*(q'_\perp)D(q_\perp)(1-x)^3x},\\
    H_{1,4}&=0,\\
    H_{1,5}&=\frac{m^2N\Delta^2_\perp}{D^*(q'_\perp)D(q_\perp)x^2(k_2\Delta_1-k_1\Delta_2)},\\
    H_{1,6}&=-\frac{m^2Nk_\perp\cdot\Delta_\perp}{D^*(q'_\perp)D(q_\perp)x^2(k_2\Delta_1-k_1\Delta_2)},\\
    H_{1,7}&=-\frac{2m^2Nk_\perp\cdot\Delta_\perp}{D^*(q'_\perp)D(q_\perp)x^2(1-x)(k_2\Delta_1-k_1\Delta_2)},\\
    H_{1,8}&=\frac{2m^2Nk^2_\perp}{D^*(q'_\perp)D(q_\perp)x^2(1-x)(k_2\Delta_1-k_1\Delta_2)}.
  \end{align}
  In the limit  $\xi\rightarrow 0$, the second term of GTMDs vanishes with contribution only coming from the first term, where the factor $(x^2-\zeta^2)$ is replaced by $x^2$ in the denominator. The factor $D(q_\perp,y)D^*(q'_\perp,x')$ present in $\alpha$ also changes accordingly to $D(q_\perp,x)D^*(q'_\perp,x)$ which can be simply written as $D(q_\perp)D^*(q'_\perp)$. 
 
\section{Wigner distribution in sigma space\label{Sec_WD_sigma}}

Wigner distribution contains the information on the momentum space $(x,k_\perp)$ as well as the position equivalent spaces $\sigma, b_\perp$, and thus is useful to investigate the non-perturbative structures of a system. The transverse impact parameter $b_\perp$ is Fourier conjugate to the transverse momentum transfer $D_\perp=\Delta_\perp/(1-\xi^2)$ \cite{Diehl:2002he,Burkardt:2002hr,Ralston:2001xs,Kaur:2018ewq} and longitudinal boost invariant space variable $\sigma=\frac{1}{2}b^- P^+$ is the  Fourier conjugate to the longitudinal momentum transfer $\xi$ \cite{Brodsky:2006in,Brodsky:2006ku}. The Wigner distribution in the transverse impact parameter space $W^{\nu [\Gamma]}_{[\lambda^{\prime\prime}\lambda^{\prime}]}(x,\xi,b_\perp, p_\perp)$  can be found from the Fourier integration of GTMDs correlator of Eq.(\ref{qqc}) over $D_\perp$. Note that at $\xi=0 $ limit, $D_\perp=\Delta_\perp$. For this dressed quark system, quark and gluon Wigner Distributions in transverse impact parameter space $W^{\nu [\Gamma]}_{[\lambda^{\prime\prime}\lambda^{\prime}]}(x,\xi,b_\perp, p_\perp)$ is extensively studied in this model for $\xi=0$ limit only \cite{More:2017zqp,More:2017zqq,Mukherjee:2014nya}. In this work, we concentrate on the Wigner distribution in the boost invariant longitudinal space $W^{\nu [\Gamma]}_{[\lambda^{\prime\prime}\lambda^{\prime}]}(x,\sigma,\Delta_\perp, p_\perp)$ which is found from the Fourier integration of Eq.(\ref{qqc}) over the skewness $\xi$. Quark sector is presented in this section only and the gluon sector is beyond the scope of this work and kept reserved for the future works. Recently some works came out addressing the boost invariant longitudinal position space \cite{Miller:2019ysh,Maji:2022tog} and have attracted major attention of the community. 
The Wigner distribution in longitudinal impact parameter ($\sigma$) space is defined as 
\begin{align}
 \rho^{[\Gamma]}(x,\sigma,\Delta_\perp,k_\perp;S)&=\int_{0}^{\xi_{max}}\frac{d\xi}{2\pi}e^{i\sigma\cdot\xi}W^{[\Gamma]}(x,\xi,\Delta_\perp,k_\perp;S)
\end{align}
 where upper limit of the integration is restricted by the energy transfer $t$ to the system as 
         \begin{equation*}
          \xi_{max}=\frac{-t}{2m^2}\Big(\sqrt{1+\frac{4m^2}{-t}}-1\Big) ~;~~~~
         \text{and}~~-t=\frac{4\xi^2m^2+\Delta^2_\perp}{1-\xi^2}. \label{rel_t_Del}
        \end{equation*}
Here $S$ is the polarization of the dressed quark system state.
The different polarization projection of Wigner distributions are denoted by the symbol $\rho_{X,Y}$, where $X$ and $Y$ represent the polarization of dressed quark state and the struck quark respectively and the subscripts $X, Y = U, L, T $ for the unpolarized, longitudinally polarized and transverse polarization respectively. For the different polarization combination of $X$ and $Y$, the wigner distributions reads as
\be
\rho_{UY}(x,\sigma,\Delta_\perp,k_\perp) &=& \frac{1}{2}\Big[\rho^{[\Gamma]}(x,\sigma,\Delta_\perp,k_\perp,+\hat{e}_z) +
\rho^{[\Gamma]}(x,\sigma,\Delta_\perp,k_\perp,-\hat{e}_z) \Big] \label{rUY}\\
\rho_{LY}(x,\sigma,\Delta_\perp,k_\perp) &=& \frac{1}{2}\Big[\rho^{[\Gamma]}(x,\sigma,\Delta_\perp,k_\perp,+\hat{e}_z) -
\rho^{[\Gamma]}(x,\sigma,\Delta_\perp,k_\perp,-\hat{e}_z) \Big] \label{rLY} \\
\rho^i_{TY}(x,\sigma,\Delta_\perp,k_\perp) &=& \frac{1}{2}\Big[\rho^{[\Gamma]}(x,\sigma,\Delta_\perp,k_\perp,+\hat{e}_i) -
\rho^{[\Gamma]}(x,\sigma,\Delta_\perp,k_\perp,-\hat{e}_i) \Big]   \label{rTY}
\ee
Each of the above equations Eqs.(\ref{rUY}, \ref{rLY}, \ref{rTY}) is a composite form of three Wigner distributions corresponding to the three different polarization of quark $Y=\{U, L, T\}$ corresponding gamma structures $\Gamma = \{ \gamma^+, \gamma^+ \gamma^5, \sigma^{+j}\gamma^5\}$ respectively. In  Eqs.(\ref{rTY}), the flouting index $i=\hat{x}, \hat{y}$ represent the direction of the transversely polarized dressed quark system in the transverse plane. 
%\item \underline{The pretzelous distribution}
One can also find the pretzelocity Wigner distribution when quark and dressed quark both are transversely polarized but in the mutually orthogonal direction and defined as
\be
\rho^{\perp j}_{TT}(x,\sigma,\Delta_\perp,k_\perp) &=& \frac{1}{2}\epsilon^{ij}_{\perp}\Big[\rho^{[\sigma^{+j}\gamma^5]}(x,\sigma,\Delta_\perp,k_\perp,+\hat{e}_i)-
\rho^{[\sigma^{+j}\gamma^5]}(x,\sigma,\Delta_\perp,k_\perp,-\hat{e}_i) \Big]\nn\\
\ee
Here $\pm\hat{e}_i$ is the polarization vector corresponding to the transverse polarization of the target state and can be expressed as the superposition of the longitudinal polarization vector $+\hat{e}_z$, and $-\hat{e}_z$,
\be
|\pm\hat{e}_i\rangle = \frac{1}{\sqrt{2}}\Big(|+\hat{e}_z\rangle\pm|-\hat{e}_z\rangle\Big)
\ee
%
%\subsection{Wigner distribution as the function of Generalized TMDs}
Wigner distribution are defined as the Fourier transform of the most general quark-quark correlator of Eq.(\ref{qqc}). Using the bilinear decomposition of Eq.(\ref{Bilinear-Decomposition-F},\ref{Bilinear-Decomposition-G},\ref{Bilinear-Decomposition-H}), $\rho_{XY}$ can further be parameterized in terms of different GTMDs\cite{Maji:2022tog}. Here we list up the three of them  $\rho_{UU}, \rho_{LL}, \rho_{TT}$ and $\rho^\perp_{TT}$ in terms of GTMDs which read as
           \begin{align}
               \rho_{UU}(x,\sigma, t,k_\perp)&=\int_{0}^{\xi_{max}}\frac{d\xi}{2\pi}e^{i\sigma\cdot\xi}\frac{1}{\sqrt{1-\xi^2}}F_{1,1}\\ \label{r_UU}
                \rho_{LL}(x,\sigma, t,k_\perp)&=\int_{0}^{\xi_{max}}\frac{d\xi}{2\pi}e^{i\sigma\cdot\xi}\frac{2}{\sqrt{1-\xi^2}}G_{1,4}\\ \label{r_LL}
                 \rho^j_{TT}(x,\sigma, t,k_\perp)=&\int_{0}^{\xi_{max}}\frac{d\xi}{2\pi}e^{i\sigma\cdot\xi}\epsilon^{ij}_\perp(-1)^j\Bigg[\frac{1}{2m^2\sqrt{1-\xi^2}}( k^i_\perp\Delta^i_\perp H_{1,1}+(\Delta^i_\perp)^2H_{1,2})\nn\\
              &+\sqrt{1-\xi^2}H_{1,3}+\frac{\sqrt{1-\xi^2}}{m^2}(k^j_\perp)^2H_{1,4}+\frac{1}{m^2\sqrt{1-\xi^2}}k^j_\perp\Delta^j_\perp\nn\\
              &((1-\xi^2)H_{1,5}-\xi H_{1,7})+\frac{1}{m^2\sqrt{1-\xi^2}}(\Delta^j_\perp)^2((1-\xi^2)H_{1,6}-\xi H_{1,8})\Bigg]\\
              \rho^{\perp j}_{TT}(x,\sigma, t,k_\perp)=&\int_{0}^{\xi_{max}}\frac{d\xi}{2\pi}e^{i\sigma\cdot\xi}\epsilon^{ij}_\perp\Bigg[-\frac{1}{2m^2\sqrt{1-\xi^2}}k^i_\perp\Delta^j_\perp(  H_{1,1}-2(1-\xi^2)H_{1,5})\nn\\ 
        &-\frac{1}{2m^2\sqrt{1-\xi^2}}\Delta^i_\perp\Delta^j_\perp( H_{1,2}-2(1-\xi^2)H_{1,6}-\xi H_{1,8})\nn\\
        &+\frac{\sqrt{1-\xi^2}}{m^2}k^i_\perp k^j_\perp H_{1,4}
        +\frac{\xi}{2m^2\sqrt{1-\xi^2}}k^j_\perp\Delta^i_\perp H_{1,7}\Bigg] \label{r_TTp}
           \end{align}
          
 \begin{figure}[!htp]
\begin{minipage}[c]{1\textwidth}
\small{(a)}\includegraphics[width=7.8cm,height=6cm,clip]{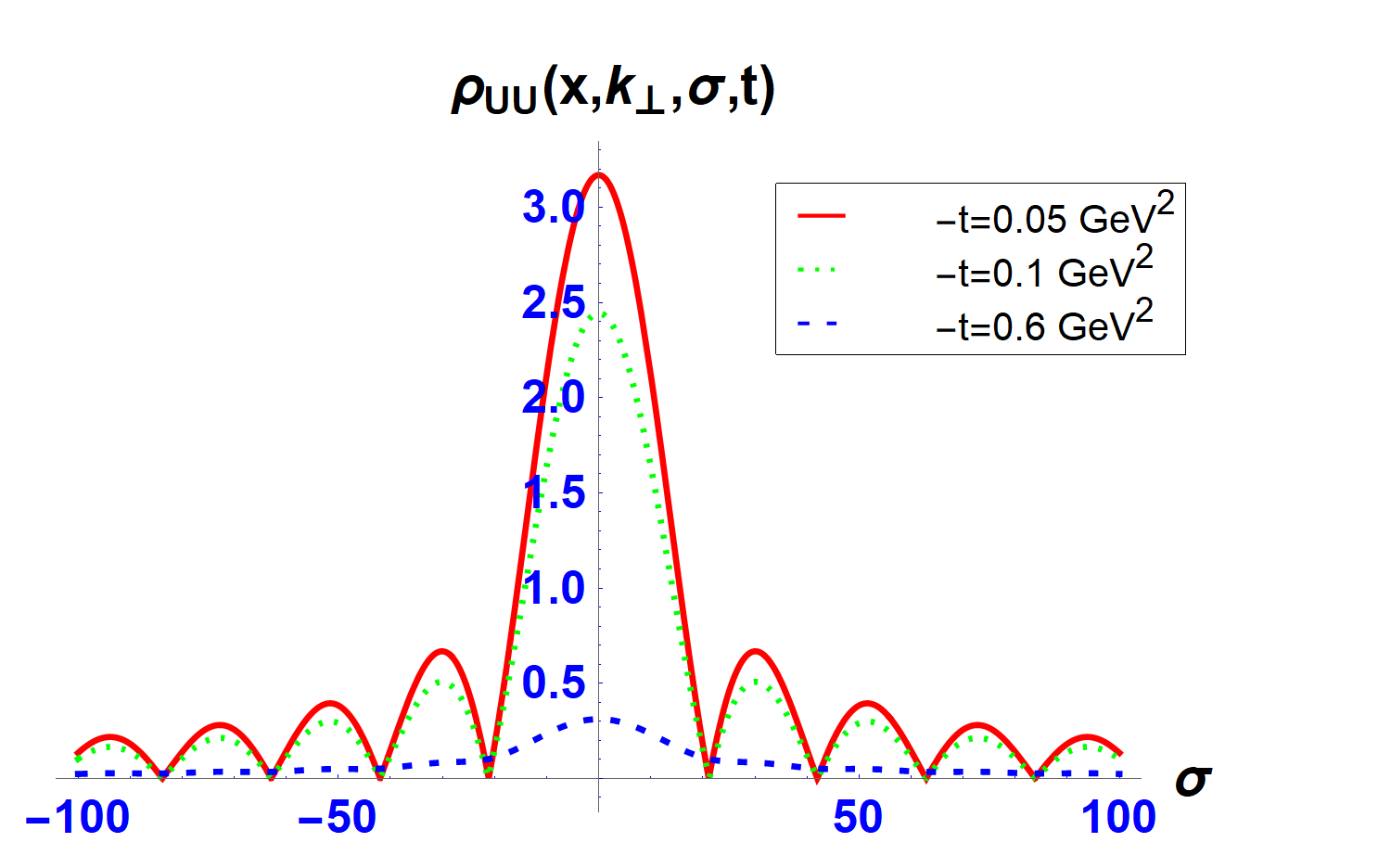}
\hspace{0.1cm}
\small{(b)}\includegraphics[width=7.8cm,height=6cm,clip]{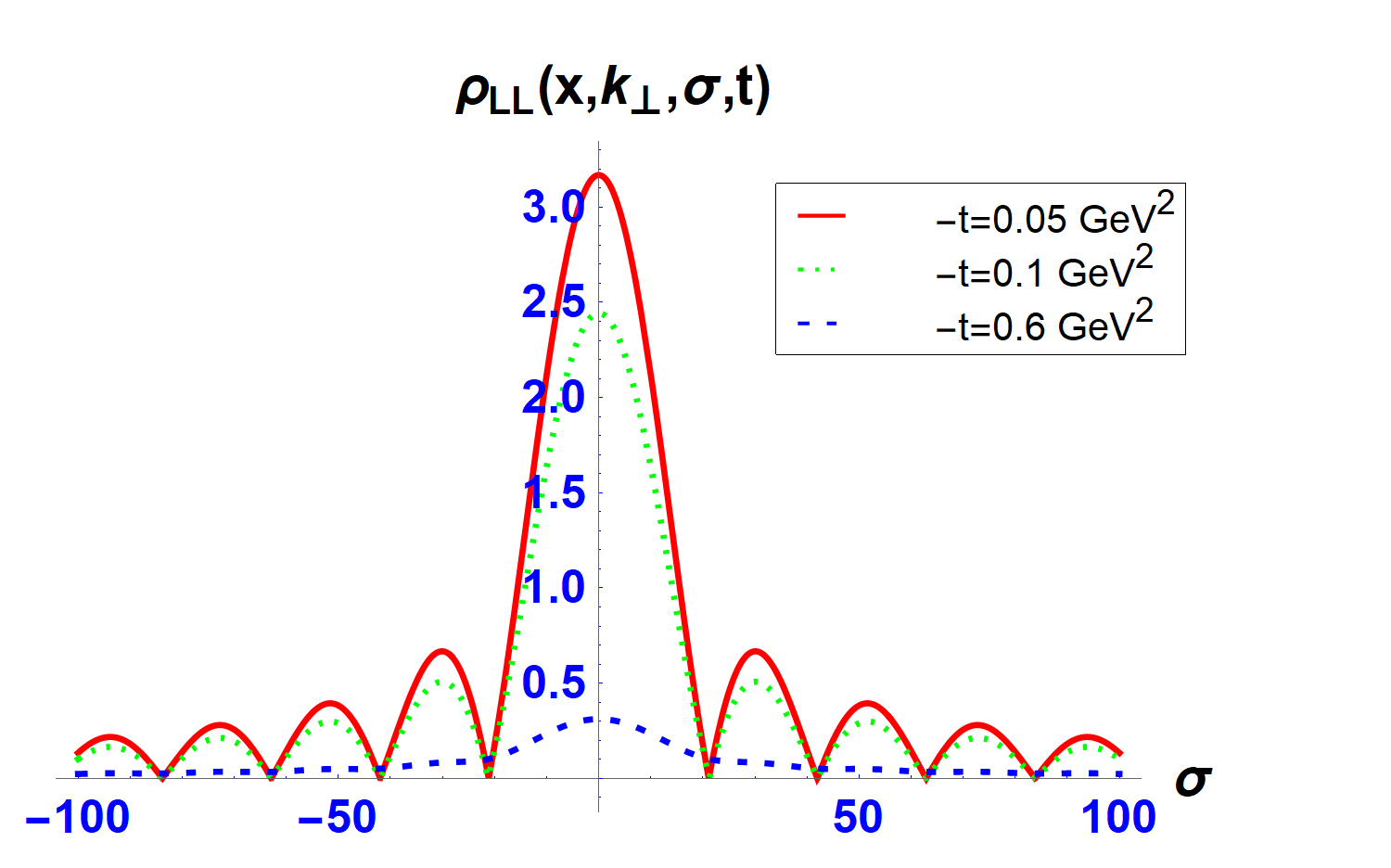} \\
\end{minipage}
\caption{\label{fig_sig}
The first moment of quark Wigner distribution in the $\sigma$-space for different values of $-t$. (a) when quark and dressed quark system both are unpolarized $\rho_{UU}$  and (b) longitudinally polarized $\rho_{LL}$.  }
\end{figure}
\begin{figure}[!htp]
\begin{minipage}[c]{1\textwidth}
\small{(a)}\includegraphics[width=7.8cm,height=6cm,clip]{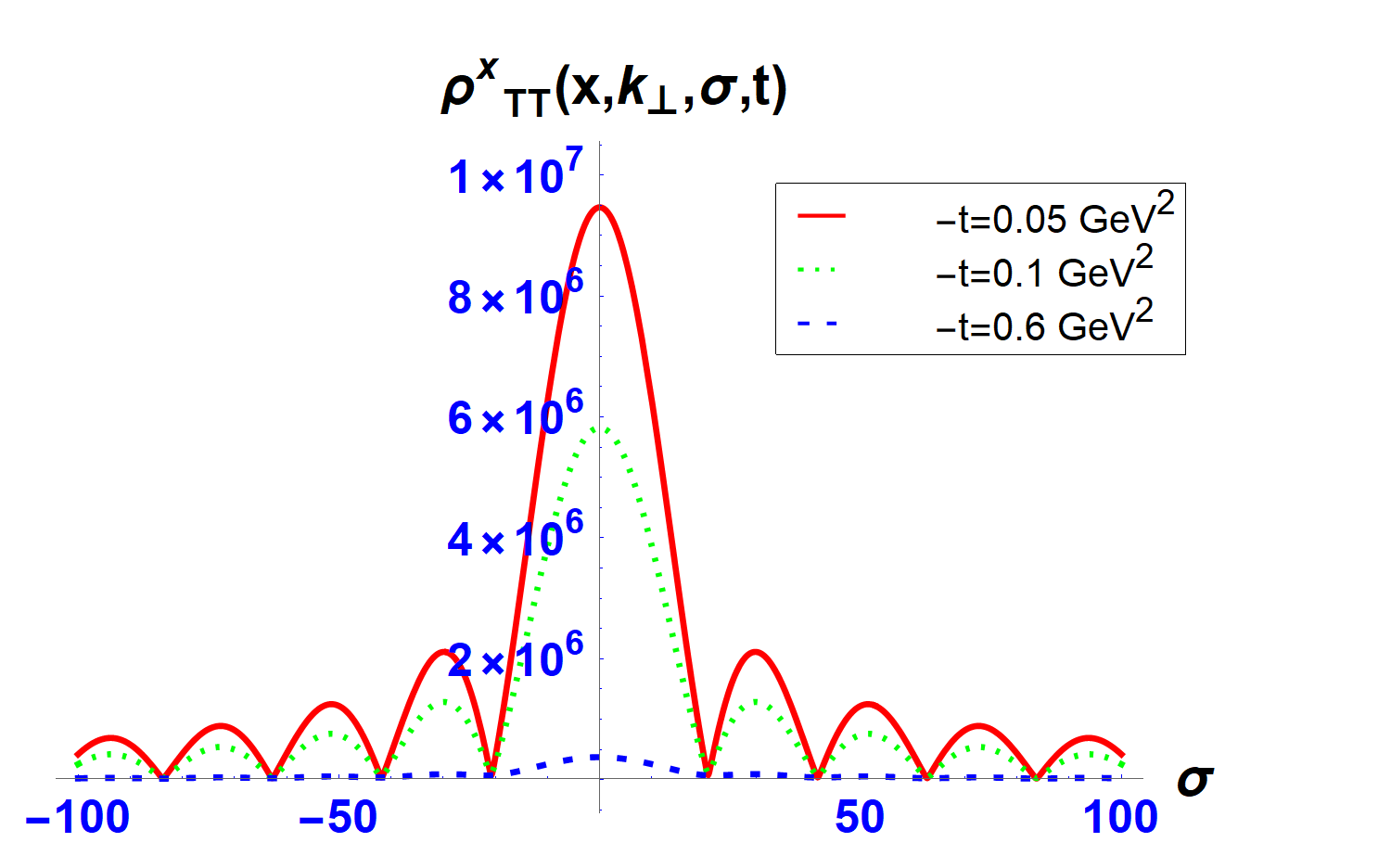}
\hspace{0.1cm}
\small{(b)}\includegraphics[width=7.8cm,height=6cm,clip]{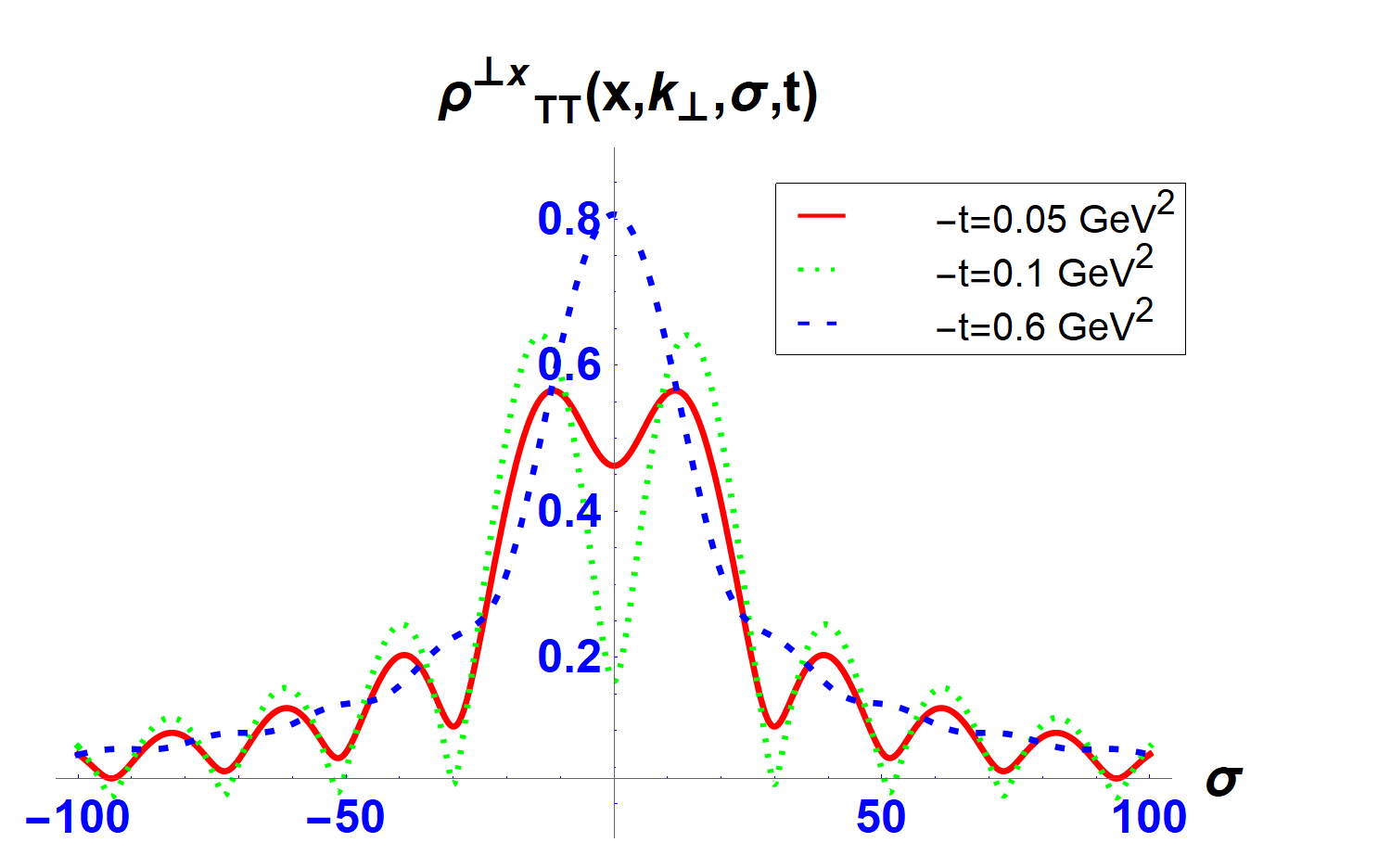} \\
\end{minipage}
\caption{\label{fig_sig2}
The first moment of quark Wigner distribution in the $\sigma$-space for different values of $-t$-- (a) when quark and dressed quark system both are transversely polarized $\rho^x_{TT}$ along $\hat{x}$,  (b) when quark and dressed quark system both are transversely polarized but they are mutually perpendicular $\rho^{\perp x}_{TT}$, struck quark is along $\hat{x}$.  }
\end{figure}

 The explicit form of the GTMDs in this dressed quark model are given in Eqs.(\ref{f11}-\ref{h18}). Using those GTMDs, the numerical results of above Wigner distributions in the $\sigma$ space are shown in Fig.\ref{fig_sig} and \ref{fig_sig2}.  Note that, the right hand side of Eqs.(\ref{r_UU}-\ref{r_TTp}) (including GTMDs) are functions of $ \chi_{1,j} (x,\sigma, \Delta_\perp, k_\perp)$ and we replace the argument $\Delta_\perp$ by $t$ using the relation between them as given in Eq.(\ref{rel_t_Del}) and then performed the Fourier transformation to $\sigma$-space. In Fig.(\ref{fig_sig}), we illustrated the distributions $\rho_{UU}$ and  $\rho_{LL}$ as a function of $\sigma$ at fixed $x=0.3$ and $k_\perp= 0.2\hat{i}$ GeV. These distributions involves $F_{1,1}$ and $G_{1,1}$  when both struck quark and dressed quark system are unpolarized and longitudinally polarized respectively. The three plots in each sub-figures are for different values of $-t={0.05, 0.1, 0.6}$ GeV$^2$ and the corresponding $\xi_{max}= {0.999782, 0.999891, 0.999982}$. We chose $\Delta_\perp \perp k_\perp $ and reduces the contribution from the terms containing $\bf \Delta_\perp . k_\perp $.  
 
 The $\sigma$ variation shows a oscillatory behavior that can be considered as diffraction type pattern. This diffraction pattern is found to be similar to the single slit diffraction of lights in optics. The width and peak of the central maxima reduces with the increasing value of $-t$ and essentially the skewness plays role analogues to the slit-width in single-slit diffraction of optics. A similar pattern is reported for Wigner distribution in longitudinal boost invariant space in the other model results for different process\cite{Miller:2019ysh,Maji:2022tog}.   
 
 Fig.(\ref{fig_sig2}) represent the $\sigma$-space variations for $\rho_{TT}$ and $\rho^\perp_{TT}$ that shows a similar diffraction pattern.

%%%%%%%%%%%%%%%%%%%%%%%%%%%%%%
%\subsection{Wigner Distribution and Diffraction Pattern}
%%%%%%%%%%%%%%%%%%%%%%%%%5%%%%%

\section{Conclusion}

In this work, we present leading twist GTMDs for non-zero skewness and study Wigner distributions in the boost-invariant longitudinal space in light-front dressed quark model. Here we concentrate on the quark sectors and investigated the distributions for different polarization combination of struck quark and the dressed quark system. We obtained the analytical expression for 16 GTMDs at twist 2 for quark in the light-front dressed quark model at non-zero skewness and discussed the numerical results of few GTMDs which has physical significance at certain limit.
The obtained results for unpolarized and longitudinally polarized quark in the limit $\xi\rightarrow 0$ agrees with  \cite{Mukherjee:2014nya}, where the GTMDs has been calculated at zero skewness. At the forward limit and $\xi \to 0$, GTMDs $F_{1,2}$ and $H_{1,1}$ reduces to Sivers and Boer-Mulders function respectively.  The model result for both GTMDs $F_{1,2}$ and $H_{1,1}$ show that the distribution peak shifts towards large $x$ with the increase of $\Delta^2_\perp$. The GTMDs $F_{1,4}$ and $G_{1,1}$ provides the spin-OAM and spin-spin correlation among the struck quark and the dressed quark system. The $\xi$ dependence of GTMDs does not make any difference to the orbital angular momentum and spin-orbit correlation of quark as both the quantities are defined for the limit $\xi\rightarrow 0$. However, the $\xi$ dependent expression of GTMDs are significant to define the Wigner distribution of quarks in boost-invariant sigma space. 
A total of 16 Wigner distribution has been defined depending on the polarization of quark and dressed quark. All the Wigner distribution shows oscillatory pattern and few of them are analogous to the single slit diffraction pattern found in optics. The light-front quark-diquark model has obtained a similar result in boost-invariant longitudinal  space \cite{Maji:2022tog} for quark distribution inside a proton. This diffraction pattern is not surprising as such behavior has previously been seen for GPD in the $\sigma$-space. An essential feature of dressed quark model is that it allows for studying gluon behavior in a bound state and the gluon sector is going to present in next work.

\section*{Acknowledgment}
VKO is supported by the seed grant project, SVNIT Surat, with the assigned project number 2021-22/DOP/05. T. M.  thanks the Science and Engineering Research Board (SERB) for the support through the National postdoctoral Fellowship (NPDF) grant of file No. PDF/2021/001964. 
\newpage
\appendix

\section{GTMDs in Modified Form}

\begin{align}
      {W^{[\gamma^{+}]}_{\lambda,\lambda'}}&=\frac{1}{2m}\Bar{u}(p',\lambda')\Big[F_{1,1}-\frac{i\sigma^{i+}k_{i\perp}}{P^+}F_{1,2}-\frac{i\sigma^{i+}\Delta_{i\perp}}{P^+}F_{1,3}+\frac{i\sigma^{ij}k_{i\perp}\Delta_{j\perp}}{m^2}F_{1,4}\Big]u(p,\lambda)\label{Bilinear-Decomposition-F}\\
       W^{[\gamma^{+}\gamma_5]}_{\lambda,\lambda'}&=\frac{1}{2m}\Bar{u}(p',\lambda')\Big[\frac{-i\epsilon^{ij}_{\perp}k_{i\perp}\Delta_{j\perp}}{m^2}G_{1,1}-\frac{i\sigma^{i+}\gamma_5 k_{i\perp}}{P^+}G_{1,2}-\frac{i\sigma^{i+}\gamma_5 \Delta_{i\perp}}{P^+}G_{1,3} +i\sigma^{+-}\gamma_5 G_{1,4}\Big]u(p,\lambda) \label{Bilinear-Decomposition-G}\\
        W^{[i\sigma^{+j}\gamma_5]}_{\lambda\lambda'}&=\frac{1}{2m}\Bar{u}(p',\lambda')\Big[-\frac{i\epsilon^{ij}_\perp p^i_\perp}{m}H_{1,1}-\frac{i\epsilon^{ij}_\perp \Delta^i_\perp}{m}H_{1,2}+\frac{mi\sigma^{j+}\gamma^5}{P^+}H_{1,3}+\frac{p^j_\perp i \sigma^{k+}\gamma^5p^k_\perp}{mP^+}H_{1,4} \nn\\
        & +\frac{\Delta^j_\perp i \sigma^{k+}\gamma^5p^k_\perp}{mP^+}H_{1,5}+\frac{\Delta^j_\perp i \sigma^{k+}\gamma^5\Delta^k_\perp}{mP^+}H_{1,6}
                    +\frac{p^j_\perp i \sigma^{+-}\gamma^5}{m}H_{1,7}+\frac{\Delta^j_\perp i \sigma^{+-}\gamma^5}{m}H_{1,8}\Big]u(p,\lambda) \label{Bilinear-Decomposition-H}
  \end{align}

\begin{enumerate}[label=(\alph*)]
    \item \underline{For unpolarized quark},
           \begin{align}
               \rho_{UU}(x,\sigma,\Delta_\perp,k_\perp)&=\int_{0}^{\xi_{max}}\frac{d\xi}{2\pi}e^{i\sigma\cdot\xi}\frac{1}{\sqrt{1-\xi^2}}F_{1,1}\\
               \rho_{UL}(x,\sigma,\Delta_\perp,k_\perp)&=\int_{0}^{\xi_{max}}\frac{d\xi}{2\pi}e^{i\sigma\cdot\xi}\frac{-i}{m^2\sqrt{1-\xi^2}}\epsilon^{ij}_\perp k^i_\perp\Delta^j_\perp G_{1,1}\\
                \rho^j_{UT}(x,\sigma,\Delta_\perp,k_\perp)&=\int_{0}^{\xi_{max}}\frac{d\xi}{2\pi}e^{i\sigma\cdot\xi}\frac{-i}{m^2\sqrt{1-\xi^2}}\epsilon^{ij}_\perp\Big[k^i_\perp H_{1,1}+\Delta^i_\perp H_{1,2}\Big]
           \end{align}
    \item \underline{For longitudinal polarized quark},
           \begin{align}
\rho_{LU}(x,\sigma,\Delta_\perp,k_\perp)&=\int_{0}^{\xi_{max}}\frac{d\xi}{2\pi}e^{i\sigma\cdot\xi}\frac{i}{m^2\sqrt{1-\xi^2}}\epsilon^{ij}_\perp k^i_\perp\Delta^j_\perp F_{1,4}\\
\rho_{LL}(x,\sigma,\Delta_\perp,k_\perp)&=\int_{0}^{\xi_{max}}\frac{d\xi}{2\pi}e^{i\sigma\cdot\xi}\frac{2}{\sqrt{1-\xi^2}}G_{1,4}\\
               \rho^j_{LT}(x,\sigma,\Delta_\perp,k_\perp)&=\int_{0}^{\xi_{max}}\frac{d\xi}{2\pi}e^{i\sigma\cdot\xi}\frac{2}{m\sqrt{1-\xi^2}}\Big[k^j_\perp H_{1,7}+\Delta^j_\perp H_{1,8}\Big]
           \end{align}
    \item \underline{For transversely polarized quark},
           \begin{align}
               \rho^i_{TU}(x,\sigma,\Delta_\perp,k_\perp)=&\int_{0}^{\xi_{max}}\frac{d\xi}{2\pi}e^{i\sigma\cdot\xi}\frac{-i}{2m\sqrt{1-\xi^2}}\epsilon^{ij}_\perp\Big[\Delta^j_\perp(F_{1,1}-2(1-\xi^2)F_{1,3})\nn\\
               \rho^i_{TL}(x,\sigma,\Delta_\perp,k_\perp)=&\int_{0}^{\xi_{max}}\frac{d\xi}{2\pi}e^{i\sigma\cdot\xi}\Bigg[\frac{-1}{2m^3(1-\xi^2)^{\frac{3}{2}}}\epsilon^{ij}_\perp\epsilon^{kl}_\perp k^k_\perp\Delta^l_\perp\Delta^j_\perp G_{1,1}+\nn\\
               &\frac{\sqrt{1-\xi^2}}{m}k^i_\perp G_{1,2}+\frac{1}{m\sqrt{1-\xi^2}}\Delta^i_\perp((1-\xi^2)G_{1,3}
               -\xi G_{1,4})\Bigg]\\
              \rho^j_{TT}(x,\sigma,\Delta_\perp,k_\perp)=&\int_{0}^{\xi_{max}}\frac{d\xi}{2\pi}e^{i\sigma\cdot\xi}\epsilon^{ij}_\perp(-1)^j\Bigg[\frac{1}{2m^2\sqrt{1-\xi^2}}( k^i_\perp\Delta^i_\perp H_{1,1}+(\Delta^i_\perp)^2H_{1,2})\nn\\
              &+\sqrt{1-\xi^2}H_{1,3}+\frac{\sqrt{1-\xi^2}}{m^2}(k^j_\perp)^2H_{1,4}+\frac{1}{m^2\sqrt{1-\xi^2}}k^j_\perp\Delta^j_\perp\nn\\
              &((1-\xi^2)H_{1,5}-\xi H_{1,7})+\frac{1}{m^2\sqrt{1-\xi^2}}(\Delta^j_\perp)^2((1-\xi^2)H_{1,6}-\xi H_{1,8})\Bigg]
           \end{align}
   the pretzelocity Wigner distribution is given in Sec.\ref{Sec_WD_sigma}.

\end{enumerate}

\bibliography{RefGTMD.bib}
\bibliographystyle{unsrt}
\end{document}